# Experimental realization of a full-band wave antireflection based on temporal taper metamaterials

Haonan Hou[1], Kai Peng[1], Yangkai Wang[1], Jiarui Wang[1], Xudong Zhang[1,2], Ren Wang[1], Hao Hu[3], Jiang Xiong[1*]

1 School of Physics, University of Electronic Science and Technology of China, Chengdu 611731, China.

2 School of Systems Science and Engineering, Sun Yat-sen University, Guangzhou 510275, China.

3 Key Laboratory of Radar Imaging and Microwave Photonics, Ministry of Education, College of Electronic and Information Engineering, Nanjing University of Aeronautics and Astronautics, Nanjing 211106, China.

*e-mail: xiongjiang@uestc.edu.cn

**Abstract:** As time can be introduced as an additional degree of freedom, temporal metamaterials nowadays open up new avenues for wave control and manipulation. Among these advancements, temporal metamaterial-based antireflection coatings have recently emerged as an innovative method that inherently avoids additional spatial insertions. However, prior temporal antireflection models with finite inserted temporal transition sections that rely on the destructive interference mechanism exhibit residual periodic strong reflections at high frequencies, fundamentally limiting the achievable bandwidth. In this work, the concept of "temporal taper", the temporal counterpart of a conventional spatial taper with a nearly full-band antireflection feature and good compatibility with gradual time-varying components, has been experimentally realized. A 1D temporal metamaterial base on voltage-controlled varactors has been designed experimentally validated. The temporal taper based broadband antireflection exempts the system from spatial matching insertions, and enables agile impedance matching for various terminal loads, positioning it as a promising approach in future photonic systems.

## Introduction

Controlling wave behaviors during propagation is a critical research direction in modern wave physics and engineering[1-3]. Scattering in electromagnetics, for instance, occurs when a wave encounters the interface of two media with distinct constitutive parameters, which results in a reflected wave that propagates backward in the original medium, and a transmitted wave moving forward in the second medium, redistributing energy between them. In a large portion of practical applications such as wireless communication, wireless power transfer, energy harvesting, photoelectric detection, and solar cells, non-reflective transmission is a desirable operation paradigm. Efforts to reduce reflection, or realize antireflection, range from conventional techniques, such as the quarter-wave transformer, various types of multi-section impedance transformers, and spatial tapered lines[4], to more recent concepts and structures, including biomimetics-inspired antireflection coatings[5], subwavelength surface Mie resonators[6], and metamaterials[7-11]. However, these techniques oftentimes face limitations regarding bandwidth and incident angle, or encounter challenges in practical implementation.

In recent years, time has emerged as an additional degree of freedom for manipulating electromagnetic waves, leading to the development of temporal and spatiotemporal metamaterials. These innovations have introduced a series of novel concepts, exotic phenomena, and potential applications[12-15]. Typical examples include temporal reflection and refraction[16-21], wave pattern tailoring and engineering[22-24], magnetless nonreciprocity[25-27], and modern physics involved studies such as photonic time crystals[28,29], non-Hermitian features[30,31], and temporal Anderson localization[32,33]. In this context, antireflection and impedance matching based on time modulation have also gained significant attention very recently, for they can avoid inherent limitations of the above-mentioned spatial approaches, such as additional insertions and angle dependence. Notable progress includes the first temporal analogue of the spatial quarter-wavelength impedance transformer[34], and a few subsequent studies of multi-section impedance transformers with enhanced bandwidth[35-37].

To date, however, these concepts still remain theoretical, facing significant challenges in experimental validations, which can be categorized into two main difficulties. The first is the realization of desirable temporal boundaries. Some initially proposed temporal antireflection schemes are based on ideal step-like time-varying media with abrupt temporal boundaries[34-37]. Unfortunately, the switch-like time-varying components available today generally exhibit non-negligible response time, especially at high frequencies, making the assumption of abrupt temporal boundaries unrealistic in realistic designs. Some most recent studies based on RF switches[16] and photodiode switches[17] have been reported to reach finite response times on the order of a few nanoseconds and several hundred picoseconds, respectively. The second challenge in current time modulation electromagnetic systems lies in the inherent mixing and intertwining between the waves under examination and externally applied modulation. For instance, the dynamic microstrip transmission line is one of the typical time-varying systems at RF and microwave regimes[38]. In practice, to ensure that varactors remain in their normal operation state, the magnitude of transmitted signal applied on the microstrip trace line is usually much smaller than that of the modulation signal. This, however, makes it nearly impossible to distinguish the transmitted signal along the line.

Facing these realistic challenges, the concept of "temporal taper" becomes a most promising antireflection approach in terms of practical implementation. A series of theoretical studies have been conducted in the past few years, where the temporal reflection and transmission coefficients, time-reflected and transmitted wave fields, energy balance, etc. for tapers with different profiles were calculated and some inverse design examples were theoretically demonstrated[39]-[41]. Similar to the conventional spatial counterpart for impedance matching, the continuous temporal taper with so-called smooth or soft temporal boundaries, exhibits an almost full-band antireflection capability, apart from only an inherent singularity at DC. More importantly, it largely mitigates the limitations posed by the finite response time in systems with abrupt temporal boundaries, making it highly compatible with time-varying systems based on voltage-controlled varactors. In this work, we design a 1D temporal metamaterial, a temporal taper transmission line (TTTL), to realize the temporal taper. This TTTL utilizes a differential modulation scheme, which minimizes the influence of the relatively large modulated signal on the transmitted signal along the line, allowing for clear and accurate observation of the wave's time-domain behavior, particularly, the spatial and temporal reflection. Using the TTTL, we experimentally validate the theoretical temporal reflection spectrum and observe the significant temporal antireflection attributable to the temporal taper. Finally, we demonstrate the potential for eliminating the spatial discontinuities and enabling agile impedance matching across varying load conditions.

## Results and Discussion

**Continuous temporal taper**

We start the study from a medium with an abrupt temporal interface, as shown in Fig. 1(a). Without loss of generality, we assume that the constitutive parameters of an unbounded, nondispersive medium with negligible memory effects are spatially homogeneous, but change abruptly in time. At the temporal interface, the constitutive parameters ($\varepsilon_i$, $\mu_i$) and wave intrinsic impedance $Z_i = \sqrt{\mu_i / \varepsilon_i}$ change to ($\varepsilon_f$, $\mu_f$) and $Z_f = \sqrt{\mu_f / \varepsilon_f}$ instantly. Here, we use subscripts i and f to denote the medium and wave before and after the temporal interface, respectively. When a wave encounters this temporal boundary, the temporal discontinuity produces a forward propagating time-transmitted wave and a backward propagating time-reflected wave, as indicated by the arrows and wave symbols in the figure. It has already been shown in [15] that this process is accompanied by a broadband frequency transition, which is visually represented by the color ramp in the figure. In addition, the temporal reflection and transmission coefficient for the electric field at the temporal boundary can be obtained via temporal boundary conditions and constitutive relations, and are written as

$$R_{\text{if}} = \frac{1}{2}(\frac{\varepsilon_{\text{i}}}{\varepsilon_{\text{f}}} - \frac{\sqrt{\mu_{\text{i}}\varepsilon_{\text{i}}}}{\sqrt{\mu_{\text{f}}\varepsilon_{\text{f}}}})$$

$$T_{\text{if}} = \frac{1}{2}(\frac{\varepsilon_{\text{i}}}{\varepsilon_{\text{f}}} + \frac{\sqrt{\mu_{\text{i}}\varepsilon_{\text{i}}}}{\sqrt{\mu_{\text{f}}\varepsilon_{\text{f}}}})$$

(1)

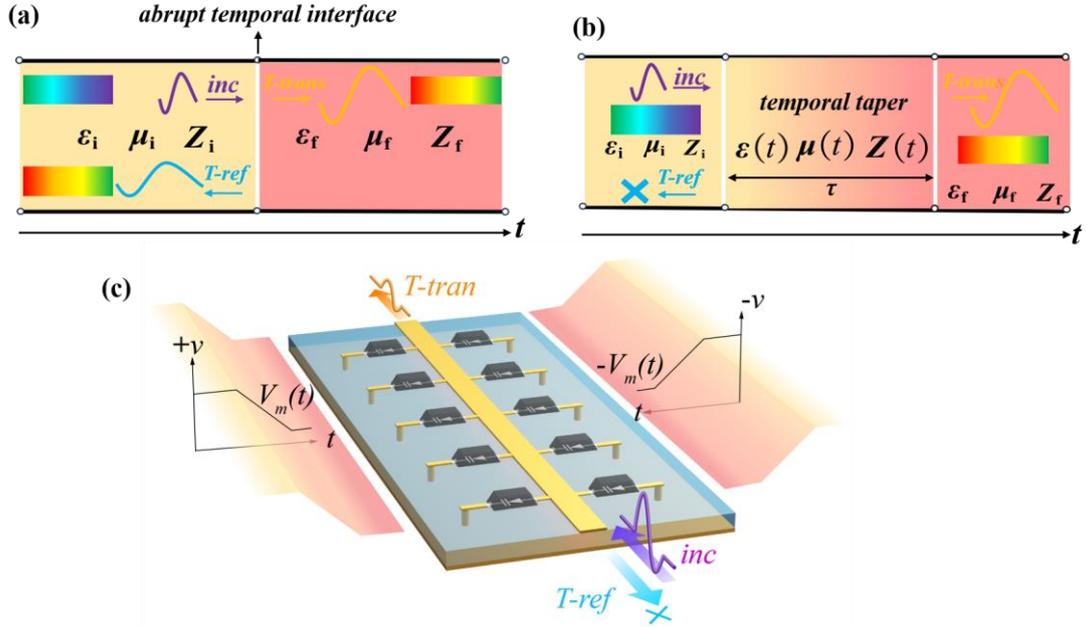

**Fig. 1. Concept and a 1D metamaterial realization of a continuous temporal taper.** (a) Abrupt temporal interface in a spatially unbounded medium showing instantaneous state switching between an initial and final state in the time-domain. ($\varepsilon_{\text{i}}$, $\mu_{\text{i}}$, $Z_{\text{i}}$) and ($\varepsilon_{\text{f}}$, $\mu_{\text{f}}$, $Z_{\text{f}}$) are the medium permittivity and permeability, and the wave intrinsic impedance in these two states, respectively. A wave incident to an abrupt temporal interface undergoes temporal reflection and transmission, accompanied by a broadband frequency transition. In the figure, the arrows, wave symbols, and color ramps denote the wave propagation directions, time-domain waveforms, and frequency spectra, respectively. (b) Continuous temporal taper with a time-varying medium constitutive parameters and wave impedance, achieving a full-band antireflection. (c) A typical 1D temporal metamaterial to implement the temporal taper in (b). The metamaterial can be a TTTL, a microstrip line with periodically loaded varactor pairs, which are under a tapered voltage modulation in the time-domain. The modulation voltage profile applied on the varactors is represented by two ramps on the two sides of the TTTL. Note the differential modulation scheme enables a clear observation of the time-domain wave behavior along the line.

In order to smooth the discontinuity between the two states and reduce the reflection, the insertions of multiple transition time segments[34-37] and temporal taper[34-37] have been suggested, respectively, both operating as a temporal impedance transformer or temporal antireflection coating. The temporal taper has time-varying constitutive parameters $\varepsilon(t)$, $\mu(t)$, and a wave impedance $Z(t)$ inserted in the time-domain between the initial and final state, which is schematically shown in Fig. 1(b). As a brief review and comparison shown in Supplementary Note 1, such a temporal taper has a much superior antireflection bandwidth than the approach of multiple transition time segments, featuring an almost full-band transmission except the intrinsic singularity at DC.

In a temporal taper, the reflection coefficient spectrum $R(\omega)$ is a key metric for evaluating the antireflection performance. While previous derivations exist, e.g. based on temporal boundary conditions and momentum conservation[39], or a Riccati's equation type formulation[40], here, we provide an alternative solution in explicit form, incorporating both amplitude and phase:

$$R(\omega) = \frac{R_{if}}{\ln(Z_f/Z_i)} \left[ \int_0^\tau \frac{Z'(t)}{Z(t)} e^{2j\int_0^t \omega \sqrt{\frac{\varepsilon_i \mu_i}{\varepsilon(t')\mu(t')}}dt'} dt \right] e^{-j\int_0^\tau \omega \sqrt{\frac{\varepsilon_i \mu_i}{\varepsilon(t)\mu(t)}}dt} \qquad (2)$$

In Eq. (2), $\varepsilon(t)$, $\mu(t)$ and $Z(t) = \sqrt{\mu(t)/\varepsilon(t)}$ are the specific forms of the time-varying permittivity, permeability and wave impedance, respectively. $Z'(t)$ is the derivative of $Z(t)$ with respect to time. $\tau$ is the total duration of the temporal taper. As for the case of only time-varying permittivity $\varepsilon(t)$ ($\mu(t)=\mu_0$) and an exponential wave impedance taper $Z(t) = Z_i e^{\frac{t}{\tau}\ln\frac{Z_f}{Z_i}}$, a typical variation adopted in a traditional spatial taper, Eq. (2) can be further reduced to

$$R(\omega) = \frac{R_{if}}{\ln(Z_f/Z_i)} \left[ \int_1^{Z_f/Z_i} \frac{e^{\frac{2j\omega\tau}{\ln(Z_f/Z_i)}t}}{t} dt \right] e^{\frac{-j\omega\tau}{\ln(Z_f/Z_i)}(Z_f/Z_i + 1)} \qquad (3)$$

Mathematical details of Eq. (2) and Eq. (3), as well as the inclusion of a non-time-varying loss, can be found in Supplementary Note 2. A validation of Equation (3) through FDTD simulations are provided in Supplementary Note 3.

**Experiment Results of Temporal Antireflection**

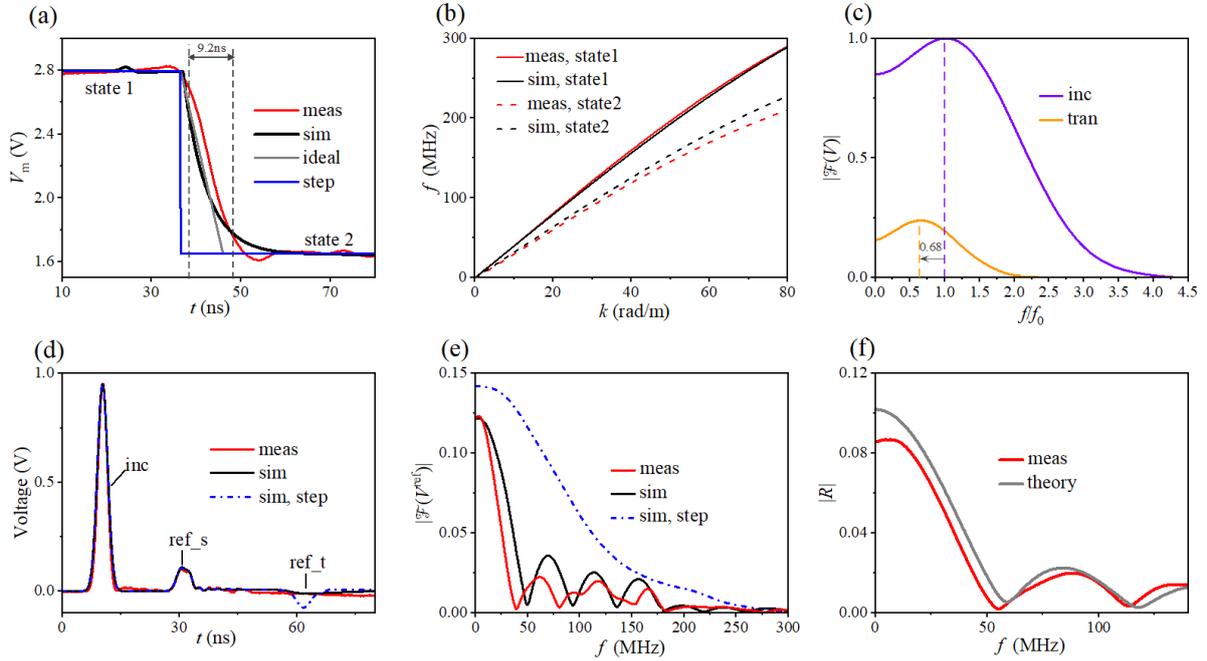

**Fig. 2. Experimental results of the temporal antireflection in a TTTL.** (a) Modulation signals applied in the simulation and experiment between two static states of the TTTL. These two states, the initial state and final state, are denoted as state 1 and 2, respectively. The grey line shows the ideal descending ramp signal as a temporal taper. The black and red lines are the corresponding simulated and measured signals on the varactors, respectively. The measured modulation voltage takes 9.2 ns to transition between 10% and 90% of the difference between states 1 and 2, indicated by the two dashed lines and a double arrow. A falling edge signal that forms a sharp transition between the two states of TTTL is plotted by the blue line for comparison. (b) Dispersion relations of the static TTTL at state 1 and 2. (c) Measured frequency translation ratio when a modulated Gaussian pulse is used as the input signal. The translation ratio for the spectrum peak is denoted by the arrow. The frequency spectrums are normalized by the carrier wave frequency of the input modulated Gaussian pulse. (d) Simulated and measured time-domain voltages at the TTTL input port. The input, spatially reflected and time-reflected waves are denoted in the figure as inc, ref_s, and ref_t, respectively. The dash-dotted line shows the case when the modulation signal is a falling edge, the blue line in (a). (e) Comparison of the frequency spectra, $|\mathscr{F}(V^{ref})|$, of the three time-reflected waves in (d). (f) Theoretical and measured amplitudes of the temporal reflection coefficient, $R(\omega) = \dfrac{\omega_f}{\omega_i} \dfrac{\mathscr{F}(V^{ref})|_{\omega_f}}{\mathscr{F}(V^{inc})|_{\omega_i}}$. Note that for a fair comparison, all simulated and measured frequency spectra in both (e) and (f) have been corrected to represent a lossless case with a compensation procedure[16].

The smoothly varying temporal taper in Fig. 1(b) is particularly suitable for time-varying systems incorporating gradual variation components, such as varactors. Therefore, a 1D temporal taper metamaterial can be designed to experimentally validate our theory. Specifically, as schematically illustrated in Fig. 1(c), the metamaterial can be implemented by a microstrip line with periodically loaded varactor pairs with a tapered voltage, which is denoted as a TTTL for brevity. According to the mapping from telegraphist's equations for transmission lines to Maxwell equations for macroscopic electromagnetic media, parameters in a transmission line circuit and the equivalent medium are quantitatively related.[42] In particular, based on such a field-circuit equivalence, a time-varying varactor capacitance $C(t)$ is equivalent to a time-varying $\varepsilon(t)$, and thus a time-varying $Z(t)$ in Fig. 1(b), if the medium has constant permeability. In particular, the two descending ramps on both sides of the microstrip line represent the modulation voltage applied on the varactors (note essential components like DC blocks are omitted in the schematic figure). Note in Fig. 1(c), the two varactors in each pair are deployed with opposite polarity. This allows their respective modulation signals cancel out and create a net-zero modulation voltage along the microstrip trace line, enabling clear observation of time-domain wave behavior (see Fig. 4(b)). We denote this as "differential modulation" scheme, which is a common microwave design technique. Detailed TTTL configuration and parameters, the differential modulation performance, as well as the experimental setup, are provided in the Methods section.

The temporal antireflection effect of the TTTL is quantitatively verified in our experiment, where a 32-unit cell TTTL in Fig. 4(c) is utilized. Fig. 2(a) shows the three stages the TTTL experiences under the time modulation in our experiment. At the beginning, the initial modulation voltage is $V_i$=2.8V. The equivalent parallel capacitance of the TTTL is accordingly $C_i$=11.47pF for each unit cell, and the TTTL is in an initial stable state, denoted as state 1 in the figure. Then, the TTTL experiences a temporal taper process. During this process, the equivalent parallel capacitance of the TTTL exponentially increases when the modulation voltage is linearly lowered down, following the descending ramp modulation signal from the AWG (see the ramp representing $V_m(t)$ in Fig. 1(c)). Finally, when the modulation voltage drops to $V_f$=1.65V, the TTTL is in the final stable state, denoted as state 2, and the TTTL equivalent capacitance arrives accordingly at its final value $C_f$=22.35pF. In Fig. 2(a), a desired ideal modulation signal is depicted by the grey line. The descending ramp signal (τ~9.2 ns) applied to the varactors was empirically optimized in the AWG output to account for the finite varactor charging time (~5ns, see Supplementary Note 4), and the actual oscillator probe detected modulation signal on the varactor (red line) exhibits slight overshoot and ringing after 50ns, which are common measurement artifacts[17]. Besides, during the experiment, we have strictly controlled both the input and modulation signal in the time-domain: when the Gaussian pulse peak reaches the TTTL midpoint, the modulation signal begins to decrease. Since the 32-unit TTTL is much longer than the input Gaussian signal in space, the transmitted signal has not yet reached the TTTL output port (the spatial boundary) even after the time variation ends. Such a situation, similar to our setup in the FDTD simulation (see Supplementary Note 3), approximates the theoretical unbounded medium assumption.

The measured results are summarized in Fig. 2(b)-(f). Firstly, The $S_{21}$ of the TTTL is measured with a two-port vector network analyzer, and it is used to obtain the static dispersion of TTTL at state 1 and 2. As shown in Fig. 2(b), the measured results align closely with the simulated ones, validating the design and fabrication accuracy of both the transmission line and loading elements. As for an input Gaussian pulse, whose main portion is at the lower frequency side (see the violet line in Fig. 2(c)), the dispersion diagram indicates that a transmitted signal with a temporal frequency below 200 MHz (corresponding to a wavenumber below 52 rad/m), is merely weakly dispersive, and therefore experiences little distortion along the TTTL.

Secondly, Fig. 2(c) shows the frequency translation ratio of the transmitted signal in the TTTL. Here a modulated Gaussian pulse with a carrier frequency of $f_0$=80MHz (the frequency spectrum peak) is particularly used, as the frequency translation phenomenon can be conveniently observed from the spectrum peak shift. The solid violet and golden lines are the normalized frequency spectrum of the input Gaussian pulse signal, and the transmitted signal at the TTTL output port after the temporal taper, respectively. The arrow shows a measured frequency translation ratio of $\frac{\omega_f}{\omega_i} = 0.68$. On the other hand, based on the aforementioned field-circuit equivalence, the equivalent constitutive parameters of TTTL in the initial ($\varepsilon_i$, $\mu_i$) and final ($\varepsilon_f$, $\mu_f$) states can be first calculated, and a theoretical frequency translation ratio can be obtained according to the momentum conservation $\omega_i \sqrt{\varepsilon_i \mu_i} = \omega_f \sqrt{\varepsilon_f \mu_f}$. Such a theoretical frequency translation ratio is $\frac{\omega_f}{\omega_i} = 0.72$, which is very close to the measured value.

Next, Fig. 2(d) exhibits the temporal antireflection effect. Firstly, when an ideal falling edge, as shown by the solid blue line in Fig. 2(a), is used as the modulation signal (no temporal taper), the detected signal at the TTTL input port is depicted as the dash-dotted blue line in Fig. 2(d). In the figure, one can easily identify the input signal around 10 ns, the reflected signal from spatial reflection around 30 ns, and the temporal reflection around 60 ns. Here the spatial reflection is due to the impedance mismatch at the input port, where the Bloch impedance of TTTL at state 1 is $Z_i=70\Omega$, and the characteristic impedance of the feeding microstrip line is 50 $\Omega$. Note the TTTL Bloch impedance at state 2 (the final state) is $Z_f=50\Omega$, which is identical to that of the feeding line at the output port, and therefore no spatially reflected wave at a later time is observed. Then, when a temporal taper, the descending ramp signal shown with the solid red line in Fig. 2(a) is applied, the temporal reflection is almost eliminated. This measured result is also in good agreement with the simulated one in Keysight Advanced Design Systems, which is depicted with the black solid line in the figure. Note that energy conservation is upheld in this time-varying system when accounting for the external modulation source. In the present configuration, the energy extracted by the source corresponds to the energy variation stored in both the varactors and the electromagnetic field during the temporal tapering process. A detailed discussion is provided in Supplementary Note 6. Fig. 2(e) shows the corresponding frequency spectra of the time-reflected waves in (d), denoted as $|\mathscr{F}(V^{ref})|$. The measured spectrum of our TTTL, greatly reduced in amplitude compare to the step-like temporal interface counterpart, generally agrees well with the simulation, illustrating a nearly full-band antireflection behavior. The prominent reflection near DC is due to the finite duration of the temporal taper, as discussed in Supplementary Note 1.

Finally, Fig. 2(f) is a comparison of the theoretical and measured reflection coefficient amplitude, the latter being calculated by $R(\omega)=\dfrac{\omega_f}{\omega_i}\dfrac{\mathscr{F}(V^{ref})|_{\omega_f}}{\mathscr{F}(V^{inc})|_{\omega_i}}$ [16]. The theoretical solution is calculated by substituting both the TTTL Bloch impedance at state 1 and 2, and the effective $\tau$ of 9.2 ns in the experiment (see the double arrow in Fig. 2(a)) into Eq. (3). Here, as a common practice, the effective $\tau$ is adopted as the duration time, corresponding to 10%~90% of the difference between the modulation voltage for state 1 and 2[17]. To ensure a fair comparison, the experimental $R(\omega)$ in Fig. 2(f), as well as the three $|\mathscr{F}(V^{ref})|$ in Fig. 2(e), has been corrected to represent a lossless case, where transmission loss along the line and power division at the T-junction have been compensated, following a procedure similar to that in [16]. Note in our experiment, the varactor inner resistance $R_s$, according to the Skyworks SMV1249 Data sheet, remains relatively stable around 2$\Omega$, with only very small fluctuations across the descending ramp modulation range of 1.6V~2.8V. Such a tiny variation is negligible compared to other non-time-varying loss sources, such as metallic and dielectric loss in the microstrip line and inevitable dissipation in $R_b$ (see Supplementary Note 4). This justifies the constant loss assumption adopted in our theory (see Supplementary Note 2). Fig. 2(f) indicates that both measured and theoretical reflection amplitudes follow the same tendency as that in a hypothetic homogeneous time-varying medium (cf. Supplementary Figure 1(b), and the FDTD simulation results in Supplementary Figure 3(c)). The small deviation between measured and theoretical results is attributed to the actual temporal taper affected by the finite charging time, as well as some other practical factors such as circuit parasitic parameters that are difficult to be quantitatively included in the theory. It is worth noting that, Fig. 2(f) presents the measured reflection coefficient only up to 140MHz. At higher frequencies, where the theoretical reflection tends to diminish, the experimentally detected spectrum reduces to a vanishing level, which is comparable to the small system noise and renders unreliable results.

Before closing this section, we emphasize that in our TTTL, the monotonic tapering modulation and non-resonant configuration ensure the system stability. Nevertheless, in more general settings, particularly in high-$Q$ resonators or under periodic temporal modulation, such as in photonic time crystals[28, 29], caution is warranted, as parametric processes may give rise to exponential growth and system instability.

**Agile Impedance Matching**

As suggested in the previous study[34], an antireflection temporal coating can be applied as an impedance transformer, offering the advantage of eliminating additional spatial insertions. Here, we show that the proposed temporal taper exhibits an additional merit: the greater flexibility for accommodating different terminal loads. The concept of the temporal impedance matching is depicted in the inset of Fig. 3(a). Suppose the initial impedance of the TTTL is $Z_i$, and a spatial boundary and discontinuity exists when the TTTL is terminated by a load with an arbitrary impedance $Z_L \neq Z_i$. Thus, when a temporal taper for the TTTL impedance is applied so that $Z_i$ gradually approaches a final value of $Z_f = Z_L$ before the transmitted wave reaches the load, an impedance transformation with vanishing reflection can be expected.

Fig. 3 shows the validation results, where the previous TTTL is applied. In our experiment, $Z_i$ is fixed to be 70 Ω, and Fig. 3(a) shows the required final modulation voltage $V_f$ for a TTTL final impedance $Z_f$, based on the varactor capacitance-voltage relationship. Here, $Z_f$ is identical to the desired terminal load impedance $Z_L$. In particular, three pairs of $V_f$ and $Z_f$ used in our experiment are indicated with square markers. For these three different $Z_L$, Fig. 3(b) plots the measured reflected wave, with and without the impedance taper. The duration of the taper is still fixed to be 9.2ns, the same as that in Fig. 2(a). Apparently, considerable spatial reflections are observed for all three cases, and the reflection is proportional to the impedance deviation ($|Z_i-Z_L|$). On the contrary, once the temporal impedance taper is applied, fairly good impedance matching can be observed and the reflections are significantly reduced. Note that such a temporal taper offers several advantages over conventional techniques such as passive spatial impedance transformers and tapers[4], as well as active non-Foster negative impedance transformers[43]. A detailed comparison is provided in the Supplementary Note 7. Although perfect impedance matching is theoretically impossible due to the presence of the intrinsic low-frequency components in the time-reflected wave spectrum, as shown in Fig. 2(e), practical improvement can be achieved by a certain extension of $\tau$ within the circuit size limitation.

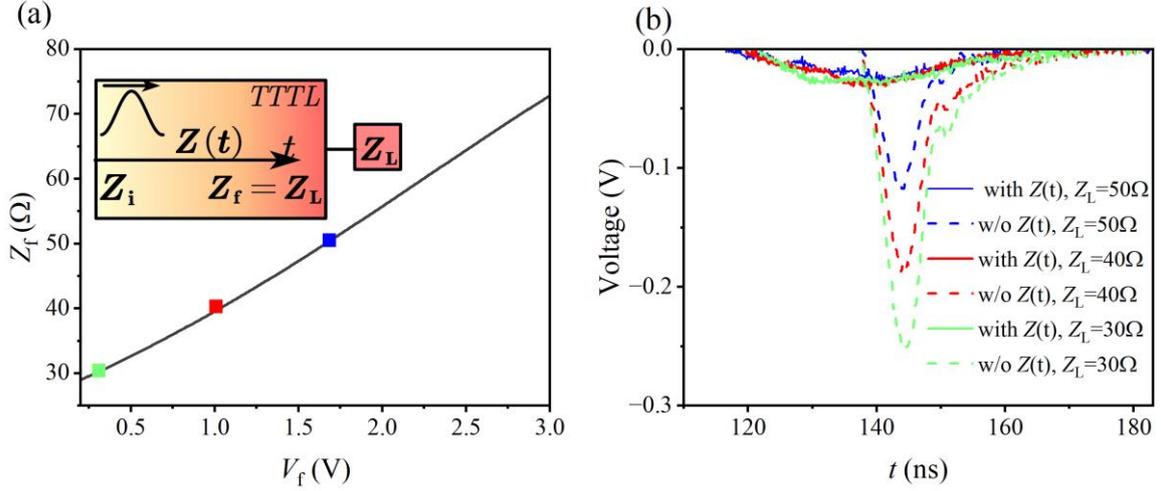

**Fig. 3. Time-varying patterns of constitutive parameters in a spatially unbounded medium with a transition from an initial state to a final state.** (a) Required modulation voltage for the final state of TTTL for an ideal impedance matching with a given load. The voltage is based on the capacitance-voltage relationship of varactors loaded on the TTTL. The cases of three different load impedances, 30Ω, 40Ω, and 50Ω, are indicated by green, red and blue square markers. The figure inset depict the schematic of temporal impedance matching using a TTTL. The impedance of TTTL gradually changes with time, from an initial value $Z_i$ to a final value $Z_f$, with the latter identical to the impedance of the load $Z_L$. (b) Measured time-reflected waves at the input ports. The initial impedance of the TTTL $Z_i$ is fixed to be 70Ω for all three cases.

As a final remark, this TTTL concept opens promising pathways for integration into various practical systems. Potential applications include reconfigurable front-ends for multi-band software-defined radio, real-time adaptive matching networks that optimize power transfer in energy harvesters or power amplifiers under varying conditions, dynamic filters for cognitive radio and frequency-hopping systems, and intelligent distributed equalizers to enhance signal integrity in high-speed data links by dynamically tailoring the transmission line dispersion profile.

## Conclusion

In summary, we have practically realized the temporal taper concept, a full-band wave antireflection approach with good compatibility to gradual time-varying components. Comparison of the frequency spectra, the TTTL periodically loaded with varactors under a differential modulation deployment, exhibits excellent temporal antireflection with frequency conversion. Although this work focuses on the exponential taper (maximum high-pass bandwidth), our approach applies directly to other profiles (e.g., triangular for minimal ripples or Klopfenstein for balanced performance) through simple modulation adjustments. Given the limited time window of the current 32-cell TTTL, which hinders clear taper profile and reflection spectra distinctions, we provide extended simulations in Supplementary Note 5 using a much longer TTTL under realistic conditions to compare these tapers effectively. Looking ahead to future practical applications, the feasibility of agile impedance matching has also been demonstrated. The temporal impedance transformer has the advantages of inherent matching capability and superior design flexibility, compared with its conventional spatial counterpart. The temporal taper can be naturally extended to higher frequency regimes, as discussed in Supplementary Note 8 regarding the response speed of key modulation components. It opens a new avenue for dynamic photonic routing and on-demand on-chip networks. For example, one promising application lies in addressing a critical challenge in integrated photonics: mitigating the modal mismatch for surface plasmon polaritons (SPPs) at the interface between different nanostructures. Similar as is demonstrated in the microwave regime, a temporal taper could dynamically manage the impedance transition, enabling cross-interface transfer with unprecedented low reflection. This work makes a considerable leap toward a practical realization for the temporal impedance matching for microwave and optical devices, while the differential modulation scheme also serves as a powerful technique for exploring novel physical phenomena with time-varying transmission lines in future.

## Methods

### Configuration of TTTL

A schematic of the experimental setup is plotted in Fig. 4(a), where the TTTL is highlighted within the green box. It is a microstrip line with periodically loaded varactors, which can be modulated by a time-varying voltage. The microstrip transmission line serves as the main path for the transmitted signal. It consists of 32-unit cells, and each unit cell is composed of a section of microstrip line, two series inductors at both ends, and two parallel differential modulation modules on both sides. Note it belongs to the symmetrical unit cell representation[42]. The series lumped inductors $L_s$ are used to reduce the transmission line wavelength and render a more compact structure. The circuit element configuration (including two critical components, $R_b$ and $C_b$) and critical design considerations for the time modulation module are illustrated in Supplementary Note 4. Meanwhile, to eliminate the influence of the relatively large modulation signal on the transmitting signal along the main transmission line, a differential modulation scheme is applied. This is the same modulation technique that we utilized in a parallel study.[44] In a differential modulation scheme, as illustrated in Fig. 4(a), a pair of varactors are loaded with opposite orientations relative to the microstrip trace line. Fig. 4(b) shows the measured performance of the differential modulation scheme. Even though the varactor modulation signal (blue lines) is larger than the input Gaussian signal (red line), modulation signals of a pair of varactors have equal amplitude but opposite phase, and therefore cancel out each other along the main path (green dashed line), ensuring precise detection of the temporal behavior of the transmitted signal.

**Experimental setup**

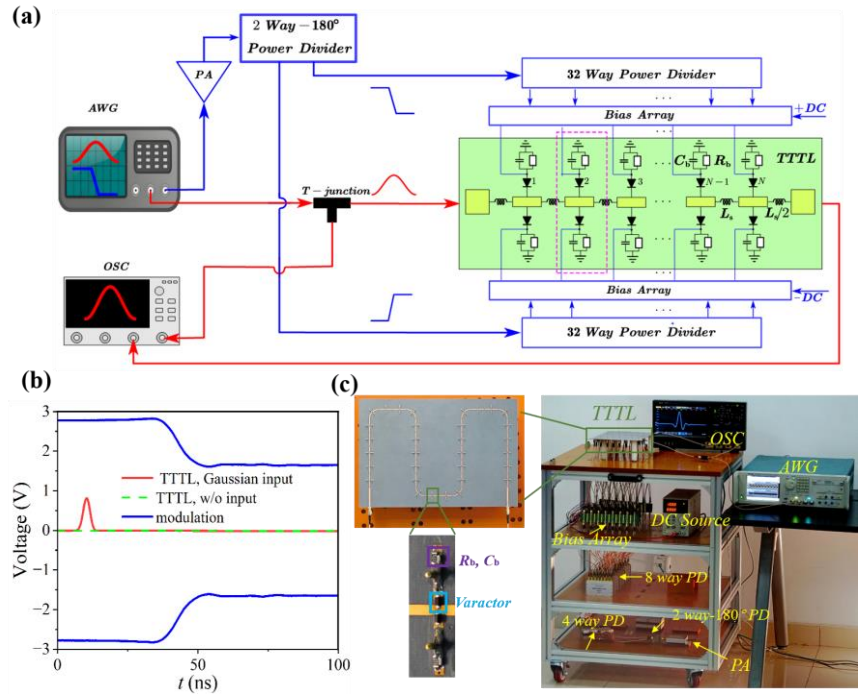

**Fig. 4. Experimental setup for the TTTL.** (a) Schematic diagram of the experimental setup. The TTTL with varactor pairs under differential modulation is highlighted in the green box. Geometrical parameters of the TTTL include: substrate relative dielectric constant of 2.2, substrate thickness of 1mm, main trace line width of 1.457 mm and two transitional section line width at two ends of 3mm (corresponding to characteristic impedances of 77 Ω and 50 Ω, respectively), and unit cell length of 27mm. The varactors

applied are Skyworks SMV1249, and other circuit elements are: $L_s$=46nH, $R_b$=51Ω, $C_b$=200pF. (b) Amplitude of the input Gaussian pulse, the measured modulation signal on a varactor pair, and the measured signal at the trace line of same TTTL unit cell without an input signal. (c) Photograph of the experimental setup and close-up for the TTTL and its unit cell. The TTTL and its modulation modules are assembled in a customized scaffold-like measurement platform[44].

The experimental setup of this work consists of the TTTL, an AWG, a digital oscilloscope and some other necessary RF components. As shown in Fig. 4(a), the dual-channel AWG simultaneously generates the Gaussian input signal and a descending ramp signal for modulation, illustrated by the red and blue arrows, respectively. As dictated by the varactor capacitance-voltage relationship, when the modulation voltage follows a linear function (the linear descending portion of the modulation signal) between an initial and final value, as shown in Fig. 2(a), the corresponding varactor capacitance follows an exponential variation, which is the desired exponential temporal taper used in Eq. (3). The Gaussian signal is split into two paths by a T-junction: one path is displayed directly on the oscilloscope, while the other is sent into the TTTL and detected by the oscilloscope at the output port. To ensure synchronous time modulation across all varactors, the modulation signal is first amplified, and separated into two differential signals by a 180° Wilkinson power divider. These two-path signals are subsequently routed through a 32-way power divider, constructed by cascading a four-way and an eight-way power divider, and are applied to the varactors together with the DC bias. Fig. 4(c) shows the fabricated TTTL, its unit cell detail, and the experimental scenario.

As is well known, system noise is a critical factor that can limit SNR or dynamic range in practical implementations. In the current setup, potential noise sources include driver noise coupling, modulation-induced spectral folding, and impedance jitter, among others. Owing to the differential modulation scheme and high-performance AWG and power amplifiers applied, a time- and spectral-domain noise measurement delivers an RMS noise voltage of 0.5 mV. This performance ensures robust observation and analysis of temporal reflection—the key phenomenon investigated in this work.

## Data Availability

The data that support the findings of this study are available from the corresponding author upon reasonable request.

**Acknowledgements**

The authors thank Mr. Mingyu Zhao (Kangxi communication technology (Shanghai) Co., Ltd.), Mr. Lidong Huang (Ruijie Networks, Chengdu), Dr. Yangyang Peng (Smarter Microelectronics Co., Ltd., Guangzhou) and Mr. Yifan Xiong (Vivo Mobile Communications Co. Ltd., Dongguan) for their helpful discussions and useful suggestions regarding the engineering aspects and application prospects of the proposed temporal taper. This work was financially supported by National Natural Science Foundation of China (U2341207).